\documentclass[preprint,eqsecnum,aps]{revtex4}

\usepackage{graphicx}%
\usepackage{dcolumn}
\usepackage{amsmath}
\usepackage{epsfig}
\usepackage{psfig}

\makeatletter
\def\btt#1{\texttt{\@backslashchar#1}}%
\DeclareRobustCommand\bblash{\btt{\@backslashchar}}%
\makeatother

\begin{document}

\title{On the Quintessence with Abelian and Non-abelian Symmetry}

\author{Xin-zhou Li}\email{kychz@shtu.edu.cn}

\author{Jian-gang Hao}
\author{Dao-jun Liu}
\author{Xiang-hua Zhai}
\affiliation{ Shanghai United Center for Astrophysics, Shanghai
Normal University, Shanghai 200234 ,China
}%

\date{\today}

\begin{abstract}
We study the perturbations on both "radial" and "angular"
components of the quintessence with an internal abelian and
non-abelian symmetry. The properties of the perturbation on the
"radial" component depend on the specific potential of the model
and is similiar for both abelian and non-abelian case. We show
that the consine-type potential is very interesting for the
O(\textit{N}) quintessence model and also give a critical
condition of instability for the potential. While the properties
of perturbations on "angular" components depend on whether the
internal symmetry is abelian or non-abelian, which we have
discussed respectively. In the non-abelian case, the fluctuation
of the "angular" component will increase rapidly with time while
in the abelian case it will not.
\end{abstract}


\maketitle \vspace{0.4cm}

\noindent\textbf{1. Introduction}

Recent observations\cite{Bernardis, Bahcall, Bennett} show that
our Universe is flat and 73 percent of its total energy density is
resulted from "dark energy", which has a negative pressure and can
accelerate the expansion of the Universe\cite{Ratra,Caldwell}.
using recent announcement by the WMAP satellite\cite{Bennett}, we
still do not judge whether dark energy is due to an unchanging,
uniform and inert "vacuum energy" (also known as a cosmological
constant) or a dynamic cosmic field that changes with time and
varies across space (known as quintessence). Quintessence can be
considered as a spatially homogeneous scalar field that evolves
with time and many
models\cite{Steinhardt,Zlatev,Coble,Kim,Chiba,6} have been
constructed so far. The possibility of quintessence as rolling
tachyon has also been investigated \cite{li}.

Boyle et al\cite{Boyle} and Gu and Hwang\cite{Gu} have discussed
quintessence with complex scalar field. In a previous paper
\cite{our}, we have further generalized their ideas by replacing
the complex scalar field with a \textit{N}-plet scalar field which
is spinning in a O(\textit{N})-symmetric potential. When one of
the angular components is fixed, this O(\textit{N}) quintessence
model will reduce to the O(\textit{N}-1) quintessence model. If
the \textit{N}-2 angular components are fixed, O(\textit{N})
quintessence model can reduce to the complex scalar model
mentioned above. If all angular components are fixed,
O(\textit{N}) quintessence model will reduce to the quintessence.

It is worth noting that this generalization does not hold its
importance for the widely studied tracker-type potentials because
the amplitude of the scalar field will increase steadily and make
the angular contribution negligible. But for another type widely
investigated potential, the cosine-type potential\cite{Kim,our},
it will be very interesting. It has been pointed out that a
natural introduction of quintessence is an ultralight axion with
an almost massless quark\cite{Kim} and the cosine-type potential
can be derived from such a model. Furthermore, This potential
requires that the amplitude of the field should not goes to be
very large and therefore, unlike the tracker-type potential, make
the angular contribution significant.

It would be very interesting to study the behaviours of the field
when it is perturbed. We firstly investigate the fluctuation of
the "radial" component and find that its stability depends on the
specific potential of the model while has nothing to do with
whether the the internal symmetry is abelian and non-abelian. We
give a critical condition of the instability for the "radial"
perturbation, under which the "radial" fluctuation will grow
rapidly with time. Fortunately, we find that this condition is not
satisfed by most potentials for quintessence models. While the
fluctuation of the "angular" components depend heavily on whether
the internal symmetry is abelian or non-abelian. In the abelian
case ($N=2$), the "angular" fluctuation will always damp, but in
the non-abelian case($N>2$), there are possibilities for the
"angular" perturbations to grow rapidly with time and become
strongly space-dependent. It is worth noting that when we take the
metric fluctuations(which are generally small compared with the
fluctuations of the quintessence field) into account, the above
conclusions are still held. In order to make the discussion more
clear, in this paper we will restrict ourselves to the $ N=2$ and
$N=3$ cases which represent the abelian and nonabelian symmetry
respectively.

\vspace{0.4cm}

\noindent\textbf{2. The Model}
 \vspace{0.4cm}

We start from the flat Robertson-Walker metric

\begin{equation}
ds^{2}=dt^{2}-a^{2}(t)(dx^{2}+dy^{2}+ dz^{2})
\end{equation}

\noindent The Lagrangian density for the quintessence with
 O(3) symmetry is

\begin{equation}
L_{\Phi}=\frac{1}{2}g^{\mu\nu}(\partial_{\mu}\Phi^{a})(\partial_{\nu}\Phi^{a})-V(|\Phi^{a}|)
\end{equation}

\noindent where $\Phi^{a}$ is the component of the scalar field,
$a=1,2,3$. To make it possess a O(3) symmetry, we write it in the
following form

\begin{eqnarray}
\Phi^{1}&=&R(t)\cos\varphi_{1}(t)\hspace{4.2cm}\nonumber\\
\Phi^{2}&=&R(t)\sin\varphi_{1}(t)\cos\varphi_{2}(t)\hspace{2.85cm}\nonumber\\
\Phi^{3}&=&R(t)\sin\varphi_{1}(t)\sin\varphi_{2}(t)\hspace{1.5cm}
\end{eqnarray}

\noindent Therefore, we have $|\Phi^{a}|=R$ and assume that the
potential of the O(3) quintessence depends only on $R$. It is
clear that when we set the component $\varphi_2$ to zero, the
above O(3) system will reduce to the O(2) abelian case.

The Einstein equations and equations of motion for the scalar
fields can be written as

\begin{equation}
H^{2}=(\frac{\dot{a}}{a})^{2}=\frac{8\pi
G}{3}[\frac{1}{2}(\dot{R}^{2}+\frac{\Omega^{2}}{a^{6}R^{2}})+V(R)]
\end{equation}

\begin{equation}
(\frac{\ddot{a}}{a})=-\frac{8\pi
G}{3}[\dot{R}^{2}+\frac{\Omega^{2}}{a^{6}R^{2}}-V(R)]
\end{equation}

\begin{equation}
\ddot{R}+3H\dot{R}-\frac{\Omega^{2}}{a^{6}R^{3}}+\frac{\partial
V(R)}{\partial R}=0
\end{equation}

\begin{equation}
\ddot{\varphi_1}+(3H+2\frac{\dot{R}}{R})\dot{\varphi_1}-\sin\varphi_1
\cos\varphi_1\dot{\varphi_2}^{2}=0
\end{equation}

\begin{equation}
\ddot{\varphi_2}+(3H+2\frac{\dot{R}}{R})\dot{\varphi_2}+2\cot\varphi_1
\dot{\varphi_1}\dot{\varphi_2}=0
\end{equation}

\noindent where $H$ is Hubble parameter and the
term$\frac{\Omega^{2}}{a^{6}R^{2}}$ comes from the first integrals
of the equations of motion for the angular components(detailed
discussion see\cite{our}).

\vspace{0.4cm} \noindent\textbf{3. General Equations of Motion for
"Radial" and "Angular" Perturbation} \vspace{0.4cm}

In order to eliminate the ambiguity from gauge freedom, one has to
identify gauge invariant quantities or choose a given gauge and
perform the calculations of perturbation in that gauge. In this
paper,  we will carry out our investigation in synchronous gauge
just as Ratra and Peebles have done in Ref.\cite{Ratra}. As we
shall see, in synchronous gauge the evolution of the modes that we
consider here depend on the gravitational field only through a
gauge invariant quantity $\frac{\dot{h}}{2}$, so we can consider a
single perturbation mode conveniently in the following sections.

The line element of perturbed metric of a spatially flat FRW
spacetime is taken as

\begin{equation}
ds^2=dt^2-a^2(t)(\delta_{ij}-h_{ij})dx^{i}dx^{j}
\end{equation}

\noindent where $h_{ij}$ are the metric fluctuations and $ \mid
h_{ij}\mid<<1 $. For simplicity, we write the first-order
equations of perturbations for the $N=3$ case. Decomposing the
components of the field as follow
\begin{equation}
R(t,\textbf{x})=R(t)+\delta R(t,\textbf{x})
\end{equation}
\begin{equation}
\varphi_1(t,\textbf{x})=\varphi_1(t)+\delta\varphi_1(t,\textbf{x})
\end{equation}
\begin{equation}
\varphi_2(t,\textbf{x})=\varphi_2(t)+\delta\varphi_2(t,\textbf{x})
\end{equation}
We can obtain the equations of motion for the fluctuations as

\begin{eqnarray}\label{generalraidal}
&&\ddot{(\delta R)}-\frac{1}{a^2}\nabla^2(\delta
R)-[\dot{\varphi_1}^2+\sin^2\varphi_1](\delta R)
+3\frac{\dot{a}}{a}\dot{(\delta R)}\nonumber\\&&+V^{''}(R)(\delta
R)-\frac{1}{2}\dot{h}\dot{R}-2\dot{\varphi_1}R(\dot{\delta
\varphi_1})\nonumber\\
&&-R\sin(2\varphi_1)\dot{\varphi_2}^2(\delta
\varphi_1)-2R\sin^2(\varphi_1)\dot{\varphi_2}(\dot{\delta
\varphi_2})=0
\end{eqnarray}

\begin{eqnarray}\label{ga}
&&\ddot{(\delta \varphi_1)}-\frac{1}{a^2}\nabla^2(\delta
\varphi_1)+3\frac{\dot{a}}{a}\dot{(\delta
\varphi_1)}+2\frac{\dot{R}}{R}(\dot{\delta \varphi_1})\nonumber\\
&&-\dot{\varphi_2}^2\cos(2\varphi_1)(\delta
\varphi_1)-\frac{1}{2}\dot{h}\dot{\varphi_1}
-\sin(2\varphi_1)\dot{\varphi_2}(\dot{\delta \varphi_2})\nonumber\\
&&+\frac{2}{R}\dot{\varphi_1}(\dot{\delta
R})-\frac{2}{R^2}\dot{R}\dot{\varphi_1}(\delta R)=0
\end{eqnarray}

\begin{eqnarray}
&&\ddot{(\delta \varphi_2)}-\frac{1}{a^2}\nabla^2(\delta
\varphi_2)+3\frac{\dot{a}}{a}\dot{(\delta
\varphi_2)}+2\frac{\dot{R}}{R}(\dot{\delta \varphi_2})-\frac{1}{2}\dot{h}\dot{\varphi_2}\nonumber\\
&&+2\cot(\varphi_1)[\dot{\varphi_1}(\delta \varphi_2)
+\dot{\varphi_2}(\dot{\delta\varphi_1})]-2\csc^2(\varphi_1)\dot{\varphi_1}\dot{\varphi_2}(\delta
\varphi_1)\nonumber\\
&&+\frac{2}{R}\dot{\varphi_2}(\dot{\delta
R})-\frac{2}{R^2}\dot{R}\dot{\varphi_2}(\delta R)=0
\end{eqnarray}

\begin{eqnarray}
\ddot{h}&+&2H\dot{h}=2\dot{R}\dot{(\delta
R)}+2R^2\dot{\varphi_1}(\delta\dot{\varphi_1})+2R^2\sin^2(\varphi_1)\dot{\varphi_2}(\dot{\delta\varphi_2})
\nonumber\\&+&2\frac{\Omega^2}{a^6R^3}(\delta
R)+R^2\sin(2\varphi_1)\dot{\varphi_2}^2(\delta\varphi_1)-V^{'}(R)(\delta
R)
\end{eqnarray}

\begin{eqnarray}
\dot{h}_{,i}-\dot{h}_{ij,j}=&&\dot{R}\partial_{i}(\delta
R)+R^2\dot{\varphi_1}\partial_{i}(\delta\varphi_1)\nonumber\\
&+&R^2\sin^2(\varphi_1)\dot{\varphi_2}\partial_{i}(\delta
\varphi_2)
\end{eqnarray}

\begin{eqnarray}\label{dhij}
\frac{1}{a^2}(h_{ij,kk}+h_{,ij}-h_{ik,jk}-h_{jk,ik})-3H\dot{h_{ij}}\nonumber\\
-H\dot{h}\delta_{ij}-\ddot{h_{ij}}=\delta_{ij}V^{'}(R)(\delta R)
\end{eqnarray}

\noindent where $\delta R$, $\delta\varphi_1$ and
$\delta\varphi_2$ are fluctuations of the " radial" and "angular"
components respectively. The above equations of motion for
fluctuations are the most general case for quintessence with O(3)
internal symmetry. When we consider only the "radial"
perturbation, we set $\varphi_1$, $\varphi_2$, $\delta\varphi_1$
and $\delta\varphi_1$ to zero. If we deal with the perturbation
on the "angular" component in the abelian case, what we need to
do is to set another "angular" component $\varphi_2$ together
with its fluctuation $\delta\varphi_2$ to zero. In the subsequent
sections, we will discuss them respectively.

\vspace{0.4cm} \noindent\textbf{4. Perturbation on the "Radial"
Components} \vspace{0.4cm}

As we have pointed out in the introduction, the property of the
perturbation on the "radial" component depend only on the
potential of the model and is independent of whether the internal
symmetry is abelian or non-abelian. In this section, we will show
this in detail. As we all know that scalar fields with an
internal symmetry are likely to produce Q balls or other
non-topological solitons, which have been studied by many
authors\cite{Kusenko, Kasuya, Coleman}. In their work, they
generally put the gravitational effects aside because the
magnitude of gravitational fluctuations induced by the
fluctuations of the scalar fields are far smaller than the
self-interaction of the scalar fields. Firstly, we, following the
previous study in this field\cite{Kusenko, Kasuya, Coleman}, also
investigate the perturbation without considering the metric
fluctuation. After this, we take the gravitational effects(metric
fluctuations)into account and carry out a similar study on the
fluctuations. The equations of motion for fluctuations of the
"radial" component are
\begin{eqnarray}
\ddot{(\delta R)}+3H\dot{(\delta R)}-\frac{1}{a^2}\nabla^2(\delta
R)+\frac{\Omega^2}{a^6R^4}(\delta R)\nonumber\\
+V^{''}(R)(\delta R)=0
\end{eqnarray}
\noindent which is obtained by setting $\delta\varphi_1$,
$\delta\varphi_2$ and $h$ in Eq.(\ref{generalraidal}) to zero. If
we choose for the fluctuation the following form:
\begin{equation}
\delta R(t,\textbf{x})=\delta R_0\exp[\alpha(
t)+i\textbf{k}\textbf{x}]
\end{equation}

\noindent then for nontrivial $\delta R_0$ , we have

\begin{equation}
\ddot{\alpha}+\dot{\alpha}^{2}+3H\dot{\alpha}
-\frac{\Omega^{2}}{a^{6}R^{4}}+\frac{k^{2}}{a^{2}}+
\frac{\partial^{2}V}{\partial R^{2}}=0
\end{equation}

\noindent Following the authors in Ref.\cite{Kusenko, Kasuya}, we
assume that $\alpha(t)$ is a slow-varying function, i.e.
$\ddot{\alpha}(t)<<{\dot{\alpha}}^2$ and $\dot{\alpha}\approx
\textrm{Const}$. In the following sections, we shall always hold
this assumption. Therefore, neglecting the $\ddot{\alpha}$ in the
above equation, we have

\begin{eqnarray}
\dot{\alpha}=\frac{1}{2}\bigg[-3H\pm\sqrt{(3H)^2-4(\frac{k^2}{a^2}-\frac{\Omega^2}{a^6R^4}+\frac{\partial^2V}{\partial
R^2})}\bigg]
\end{eqnarray}

If $\dot{\alpha}$ is real and positive, the fluctuation will grow
rapidly with time. Therefore the instability band for this
fluctuation is

\begin{equation}\label{aph1}
0<k^{2}<\frac{\Omega^{2}}{a^{4}R^{4}}-\frac{a^{2}\partial^{2}V}{\partial
R^{2} }
\end{equation}

\noindent From Eq.(\ref{aph1}), one can find that the instability
band depends on the specific potential of the quintessence.

When considering the metric fluctuations, one can obtain the
following equations of motion for the fluctuations

\begin{eqnarray}\label{dR}
\ddot{(\delta R)}+3H\dot{(\delta R)}-\frac{1}{a^2}\nabla^2(\delta
R)+\frac{\Omega^2}{a^6R^4}(\delta R)\nonumber\\
+V^{''}(R)(\delta R)-\frac{1}{2}\dot{h}\dot{R}=0
\end{eqnarray}

\begin{eqnarray}\label{eqforh}
\ddot{h}+2H\dot{h}=2\dot{R}\dot{(\delta
R)}+2\frac{\Omega^2}{a^6R^3}(\delta R)-V^{'}(R)(\delta R)
\end{eqnarray}

\begin{eqnarray}
\dot{h}_{,i}-\dot{h}_{ij,j}=\dot{R}\partial_{i}(\delta R)
\end{eqnarray}

\begin{eqnarray}\label{dhij1}
\frac{1}{a^2}(h_{ij,kk}+h_{,ij}-h_{ik,jk}-h_{jk,ik})-3H\dot{h_{ij}}\nonumber\\
-H\dot{h}\delta_{ij}-\ddot{h_{ij}}=\delta_{ij}V^{'}(R)(\delta R)
\end{eqnarray}

\noindent If we choose $\Omega = 0$, i.e. in the case of $N = 1$
the Eqs.(\ref{dR})-(\ref{dhij1}) will reduce to Ratra-Peebles's
results in the absence of baryonic term\cite{Ratra}. Clearly,
since the equations of motion for the metric and scalar
fluctuations(Eq.(\ref{eqforh}) and Eq.(\ref{dR})) are linear
equations, the fluctuations could be taken as the following form

\begin{equation}
\delta R(t,\textbf{x})=\delta R_0\exp[\alpha(
t)+i\textbf{k}\textbf{x}]
\end{equation}

\begin{equation}
h(t,\textbf{x})=h_0\exp[\alpha(t)+i\textbf{k}\textbf{x}]
\end{equation}

\noindent Since there are no $\nabla^2h$ term in
Eq.(\ref{eqforh}), $h$ can not oscillate rapidly and in fact, $h$
will not be able to react, in lowest order, to the rapidly
oscillating source terms in Eq.(\ref{eqforh})\cite{Ratra}. Then
for nontrivial $\delta R_0$ and $h_0$, we have

\begin{equation}
\dot{\alpha}^{2}+\bigg(3H-\frac{1}{2}\frac{h_0}{\delta
R_0}\dot{R}\bigg)\dot{\alpha}
-\frac{\Omega^{2}}{a^{6}R^{4}}+\frac{k^{2}}{a^{2}}+
\frac{\partial^{2}V}{\partial R^{2}}=0
\end{equation}

\noindent So
\begin{eqnarray}\label{cond1}
\dot{\alpha}&&=\frac{1}{2}\bigg[-(3H-\frac{1}{2}\frac{h_0}{\delta
R_0}\dot{R})\\\nonumber
&&\pm\sqrt{(3H-\frac{1}{2}\frac{h_0}{\delta
R_0}\dot{R})^2-4(\frac{k^2}{a^2}-\frac{\Omega^2}{a^6R^4}+\frac{\partial^2V}{\partial
R^2})}\bigg]
\end{eqnarray}

\noindent From Eq.(\ref{cond1}), it is clear that the instability
band is the same as that when we did not consider the metric
fluctuations. This is reasonable in physics because the metric
fluctuations induced by the fluctuations of the field are far
smaller than the fluctuations of the field and its back-reaction
to the field would be even smaller and thus negligible. So, when
considering the metric fluctuations, there should not be a
substantial change on the properties of the fluctuation of the
quintessence field.

\vspace{0.4cm} \noindent\textbf{5. Perturbation on the "Angular"
Components in the Abelian Case} \vspace{0.4cm}

In this section, we investigate the perturbation on "angular"
component in the abelian case, that is, the case in which the
quintessence fields possess a O(2) internal symmetry.

By setting $\delta R$, $h$, $\delta\varphi_2$ and $\varphi_2$ in
Eq.(\ref{ga}) to zero, we can obtain the equation of motion for
the fluctuation of "angular" component up to the first order as:

\begin{equation}\label{motion1}
\ddot{\delta\varphi_1}+(3H+2\frac{\dot{R}}{R})\dot{\delta\varphi_1}-
\frac{1}{a^2}\nabla^2\delta\varphi_1=0
\end{equation}

\noindent where $\delta\varphi_1$ is the fluctuation of the
"angular" component. If we choose
\begin{equation}\label{ansaz}
\delta \varphi_1(t,\textbf{x})=\delta
\phi_{10}\exp[\alpha(t)+i\textbf{k}\textbf{x}]
\end{equation}
\noindent then  for nontrivial $\delta\varphi_{10}$, from
Eq.(\ref{motion1}) we have

\begin{equation}
\dot{\alpha}^2+(3H+2\frac{\dot{R}}{R})\dot{\alpha}+\frac{k^2}{a^2}=0
\end{equation}

\noindent It is clear that

\begin{equation}\label{cc1}
\dot{\alpha}=\frac{1}{2}\bigg[-(3H+2\frac{\dot{R}}{R})\pm\sqrt{(3H+2\frac{\dot{R}}{R})^2
-4\frac{k^2}{a^2}}\bigg]
\end{equation}

\noindent From Eq.(\ref{cc1}), $\dot{\alpha}$ is always negative
and thus the fluctuation will damp quickly. That is, the angular
perturbation will not produce a significant "angular"
inhomogeneity and the global symmetry will not likely to become
space-dependent.

In the following, we will introduce the metric fluctuations into
our analysis. By a similar procedure as that in last section, we
can obtain the equations of motion for the fluctuations of metric
and "angular" component up to the first order as following:

\begin{eqnarray}\label{gravequ1}
\frac{1}{2}\ddot{h}+H\dot{h}=R^2\dot{\varphi_1}\dot{\delta\varphi_1}
\end{eqnarray}
\begin{eqnarray}
\dot{h}_{,i}-\dot{h}_{ij,j}=R^2\dot{\varphi_1}\partial_{i}(\delta\varphi_1)
\end{eqnarray}

\begin{eqnarray}\label{dhij}
\frac{1}{a^2}(h_{ij,kk}+h_{,ij}-h_{ik,jk}-h_{jk,ik})-3H\dot{h_{ij}}\nonumber\\
-H\dot{h}\delta_{ij}-\ddot{h_{ij}}=0
\end{eqnarray}

\begin{eqnarray}\label{gravequ2}
\ddot{\delta\varphi_1}+(3H+2\frac{\dot{R}}{R})\dot{\delta\varphi_1}-
\frac{1}{a^2}\nabla^2\delta\varphi_1-\frac{1}{2}\dot{h}\dot{\varphi_1}=0
\end{eqnarray}

Since Eq.(\ref{gravequ1})and Eq.(\ref{gravequ2}) are linear
equations, the fluctuations could be chosen in the following form

\begin{equation}\label{ansaz1}
\delta \varphi_1(t,\textbf{x})=\delta
\varphi_{10}\exp[\alpha(t)+i\textbf{k}\textbf{x}]
\end{equation}

\begin{equation}\label{ansaz2}
h(t,\textbf{x})=h_0\exp[\alpha(t)+i\textbf{k}\textbf{x}]
\end{equation}

\noindent Then for nontrivial $\delta \varphi_{10}$, we have (from
Eq.(\ref{gravequ2}))
\begin{equation}
\dot{\alpha}^2+(3H+2\frac{\dot{R}}{R}-\frac{1}{2}\frac{h_0}
{\delta\varphi_{10}}\dot{\varphi_1})\dot{\alpha}+\frac{k^2}{a^2}=0
\end{equation}

\noindent Therefore, one can obtain
\begin{eqnarray}\label{cc3} \dot{\alpha }=&&
\frac{1}{2}\bigg[-(3H+2\frac{\dot{R}}{R}-\frac{1}{2}\frac{h_0}
{\delta\varphi_{10}}\dot{\varphi_1})\\\nonumber &&\pm
\sqrt{(3H+2\frac{\dot{R}}{R}-\frac{1}{2}\frac{h_0}
{\delta\varphi_{10}}\dot{\varphi_1})^2 -4\frac{k^2}{a^2}}\bigg]
\end{eqnarray}

\noindent From Eq.({\ref{cc3}), it is clear that $\dot{\alpha}$
will always be negative even if the metric fluctuations are
considered.

\vspace{0.4cm}

\noindent\textbf{6. Perturbation on the "Angular" Components in
the Non-abelian Case } \vspace{0.4cm}

In this section, we generalize the discussions in last section to
the case in which the internal symmetry is non-abelian, that is
the symmetry group is O(3). We restrict ourselves to the case that
only one "angular" component is perturbed. This will not lose its
generality but greatly facilitate the discussion because we can
always choose a coordinate system in which the perturbation
appears in one angular direction. By setting $\delta R$, $\delta
\varphi_2$ and $h$ in Eq.(\ref{ga}) to zero, we can obtain the
equation of motion for the angular fluctuation up to the first
order as following:

\begin{equation}\label{motion2}
\ddot{\delta\varphi_1}+(3H+2\frac{\dot{R}}{R})\dot{\delta\varphi_1}-
\frac{1}{a^2}\nabla^2\delta\varphi_1-\cos2\varphi_1\dot{\varphi_2}^2\delta\varphi_1=0
\end{equation}

\noindent where $\varphi_1$ and $\varphi_2$ are the homogeneous
parts of the "angular" components. If we choose for
$\delta\varphi_1$ the same form as in Eq.(\ref{ansaz1}), then  for
nontrivial $\delta\varphi_{10}$, from Eq.(\ref{motion2})we have

\begin{equation}
\dot{\alpha}^2+(3H+2\frac{\dot{R}}{R})\dot{\alpha}+\frac{k^2}{a^2}-\cos2\varphi_1\dot{\varphi_2}^2=0
\end{equation}

\noindent It is clear that

\begin{equation}\label{cc2}
\dot{\alpha}=\frac{1}{2}\bigg[-(3H+2\frac{\dot{R}}{R})\pm\sqrt{(3H+2\frac{\dot{R}}{R})^2
-4\bigg(\frac{k^2}{a^2}-\cos2\varphi_1\dot{\varphi_2}^2\bigg)}\bigg]
\end{equation}

\noindent From Eq.(\ref{cc2}), one can find that $\dot{\alpha}$
could be positive if
\begin{equation}\label{stablecon1}
\frac{k^2}{a^2}-\cos2\varphi_1\dot{\varphi_2}^2<0
\end{equation}

\noindent That is, under the above
condition(Eq.(\ref{stablecon1})) the "angular" inhomogeneity
might grow rapidly with time and make the symmetry group become
space-dependent.

When taking into account the metric fluctuation and following a
similar process as that in section 5, we can obtain the equations
of motion for the fluctuations of "angular" component as follow

\begin{eqnarray}\label{e2}
\ddot{\delta\varphi_1}+(3H&+&2\frac{\dot{R}}{R})\dot{\delta\varphi_1}-
\frac{1}{a^2}\nabla^2\delta\varphi_1\nonumber\\&-&\cos2\varphi_1\dot{\varphi_2}^2\delta\varphi_1
-\frac{1}{2}\dot{h}\dot{\varphi_1}=0
\end{eqnarray}

The equations of motion for metric fluctuations are

\begin{eqnarray}
\ddot{h}&+&2H\dot{h}=2R^2\dot{\varphi_1}(\delta\dot{\varphi_1})+R^2\sin(2\varphi_1)\dot{\varphi_2}^2(\delta\varphi_1)
\end{eqnarray}

\begin{eqnarray}
\dot{h}_{,i}-\dot{h}_{ij,j}=R^2\dot{\varphi_1}\partial_{i}(\delta\varphi_1)
\end{eqnarray}

\begin{eqnarray}
\frac{1}{a^2}(h_{ij,kk}+h_{,ij}-h_{ik,jk}-h_{jk,ik})-3H\dot{h_{ij}}\nonumber\\
-H\dot{h}\delta_{ij}-\ddot{h_{ij}}=0
\end{eqnarray}

Similarly, we choose the fluctuations to be the form of
Eq.(\ref{ansaz1}) and Eq.(\ref{ansaz2}) and for nontrivial
$\delta\varphi_{10}$, from Eq.(\ref{e2}) we have

\begin{equation}
\dot{\alpha}^2+(3H+2\frac{\dot{R}}{R}-\frac{1}{2}\frac{h_0}
{\delta\varphi_{10}}\dot{\varphi_1})\dot{\alpha}-\cos2\varphi_1\dot{\varphi_2}^2
+\frac{k^2}{a^2}=0
\end{equation}
\noindent and therefore
\begin{eqnarray}\label{eqmotion2}
&&\dot{\alpha}=\frac{1}{2}\bigg[-(3H+2\frac{\dot{R}}{R}-\frac{1}{2}\frac{h_0}
{\delta\varphi_{10}}\dot{\varphi_1})\pm\nonumber\\&&\sqrt{(3H+2\frac{\dot{R}}{R}-\frac{1}{2}\frac{h_0}
{\delta\varphi_{10}}\dot{\varphi_1})^2
-4\bigg(\frac{k^2}{a^2}-\cos2\varphi_1\dot{\varphi_2}^2\bigg)}\bigg]
\end{eqnarray}

\noindent From Eq.(\ref{eqmotion2}), one can easily identify that
the condition for positive $\dot{\alpha}$ is the same as that in
Eq.(\ref{stablecon1}). This shows that in the O(3) case, the
instability condition for the "angular" fluctuations won't be
changed when considering the metric fluctuation, just as that we
have proved in the O(2) case.

 \vspace{0.4cm}

 \noindent\textbf{7. Conclusion And Discussion}

 \vspace{0.4cm}
In this paper, we investigate the perturbation on both "radial"
and "angular" components of the quintessence fields with an
internal abelian and non-abelian symmetry. We find that the
fluctuation of the "radial" component depends on the specific
potential of the quintessence model. Under certain condition, this
fluctuation could grow rapidly and thus make the "radial"
component space-dependent. For the "angular" perturbation, we find
that the properties of the fluctuations depend on whether the
internal symmetry group is abelian or non-abelian. In the abelian
case, the fluctuation of the "angular" component will damp rapidly
with time and therefore the symmetry group will not become
space-dependent. In the case that the internal symmetry is
non-abelian, we find that under certain condition, the "angular"
inhomogeneities might increase rapidly with time and make the
symmetry group space-dependent, or local. Here we briefly
interpret the physical meaning of this instability: $\delta
\varphi_{1}$ represents a small internal angular inhomogenity;
This inhomogenity will increase rapidly under the
condition(Eq.(\ref{stablecon1})). Note that this angular
inhomogenity is \textit{not} that of space-time. It is surely
interesting to study the quintessence with a "local" internal
symmetry, which we will investigate in a preparing work. When
taking into account the metric fluctuations induced by the
fluctuations of the quintessence field and assuming that the
back-reaction of the metric fluctuations on the quintessence field
are negligible as shown in Ref.\cite{Ratra}, the above conclusion
still hold true.

It is worth noting that we choose the non-abelian symmetry group
as O(3) in this paper. It is not difficult to generalize the O(3)
case to O(\textit{N}) case and one may find that in the
O(\textit{N}) case, the above conclusion for the non-abelian
symmetry group still hold true even if the specific condition
under which the inhomogeneity increase will change.

\vspace{0.8cm} \noindent ACKNOWLEDGMENTS

This work was partially supported by National Nature Science
Foundation of China under Grant No. 19875016, and Foundation of
Shanghai Development for Science and Technology No. 01JC14035.


\begin{thebibliography}{99}


\bibitem {Bernardis} P. de Bernardis \textit{et al.}, \textit{Nature} {\bf 404}, 955 (2000);S. Hanany \textit{et al.} \textit{Astrophys. J.} {\bf 545}, 1 (2000).
\bibitem {Bahcall} N. Bahcall, J. P. Ostriker, S. Perlmutter and
P. J. Steinhardt, \textit{Science} {\bf 284}, 1481 (1999);S.
Perlmutter \textit{et al.}, \textit{Astrophys. J.} {\bf 517}, 565
(1999); A. G. Riess \textit{et al.}, \textit{Astron. J.} {\bf
116}, 1009 (1998), astro-ph/9805201.
\bibitem {Bennett} C. L. Bennett \textit{et al.},
[astro=ph/0302207]; D. N. Spergel \textit{et al.},
[astro-ph/0302209]; G. Hinshaw \textit{et al.},[astro-ph/0302217].

\bibitem {Ratra} B. Ratra and P. J. Peebles, \textit{Phys. Rev.} {\bf D37}, 3406 (1988).
\bibitem {Caldwell} R. R. Caldwell, R. Dave and P. J. Steinhardt, \textit{Phys. Rev. Lett.} {\bf 80}, 1582 (1998).
\bibitem {Steinhardt} P. J. Steinhardt, L . Wang and I . Zlatev, \textit{Phys. Rev.} {\bf D59}, 123504 (1999), astro-ph/9812313.
\bibitem {Zlatev}I. Zlatev, L. Wang and P. J. Steinhardt, \textit{Phys. Rev. Lett.} {\bf 82},896 (1999), astro-ph/9807002.
\bibitem {Coble} K. Coble, S. Dodelson, J. Frieman, \textit{Phys. Rev.} {\bf
D55}, 1851 (1997).

\bibitem {Kim}
 J. E. Kim, \textit{JHEP} {\bf 9905}, 022 (1999);Y. Nomura, T.
Watari and T. Yanagida, \textit{Phys. Lett.} {\bf B484},
103(2000); \textit{Phys. Rev.} {\bf D61}, 105007 (2000).

\bibitem {Chiba} T. Chiba, \textit{Phys. Rev.} {\bf D64}, 103503 (2001).

\bibitem {6} A. Masiero, M. Pietroni and F. Rosati, \textit{Phys. Rev.}
{\bf D61}, 023504 (2000); T. Barreiro, E. J. Copeland and N. J.
Nunes, \textit{Phys. Rev.} {\bf D61}, 127301 (2000); E. J.
Copeland, N. J. Nunes and F. Rosati, \textit{Phys. Rev.} {\bf
D62}, 123503 (2000).

\bibitem {li}
 X. Z. Li, J. G. Hao and D. J. Liu,
 \textit{Chin. Phys. Lett.} \textbf{19}, 1584(2002);
J. G. Hao and X. Z. Li, \textit{Phys. Rev.} \textbf{D66},
087301(2002); J. S. Bagla, H. K. Jassal  and
 T. Padmanabhan,[astro-ph/0212198]; T. Padmanabhan,
[hep-th/0212290]; X. Z. Li and X. H. Zhai, Phys. Rev. \textbf{D67}
067501(2002).

\bibitem {Boyle}
L. A. Boyle, R. R. Caldwell and M. Kamionkowski, \textit{Phys.
Lett.} \textbf{B545}, 17(2002).

\bibitem {Gu}
Je-An Gu and W-Y. P. Hwang, \textit{Phys. Lett.} \textbf{B517}, 1
(2001).

\bibitem {our}
X. Z. Li, J. G. Hao and D. J. Liu,  \textit{Class. Quantum Grav.}
\textbf{19}, 6049(2002); X. Z. Li, D. J. Liu and J. G. Hao,
\textit{Chin. Phys. Lett.} \textbf{19}, 295(2002); X. Z. Li and J.
G. Hao,[hep-th/0303093].

\bibitem {Kusenko}
A. Kusenko, M. Shaposhnikov,\textit{ Phys. Lett.} {\bf B418}, 46 (1998);
S. Kasuya and M. Kawasaki, \textit{Phys. Rev.} {\bf D61}, 041301
(2000); \textit{Phys. Rev.} {\bf D62}, 023510 (2000).

\bibitem {Kasuya}
 S. Kasuya, \textit{Phys. Lett.} {\bf B515}, 121 (2001),
 astro-ph/0105408; X. Z. Li and X. H. Zhai, \textit{Phys. Lett.} {\bf
 B364}, 212 (1995); X. Z. Li, X. H. Zhai and G. Chen, \textit{Astropart. Phys.} {\bf
 13}, 245 (2000): X. Z. Li, J. G. Hao, D. J. Liu and G. Chen
, \textit{J. Phys.} \textbf{A34}, 1459 (2001); X. Z. Li and J. G.
Hao, \textit{Phys. Rev.} \textbf{D66}, 107701(2002); J. G. Hao and
X. Z. Li, \textit{Class. Quantum Grav.} \textbf{20}, 1703 (2003).

\bibitem {Coleman} S. Coleman, \textit{Nucl.
Phys.} {\bf 262}, 263 (1985).
\end{thebibliography}
\end{document}